\documentclass[preprint2]{aastex}
\slugcomment{Submitted to the Astrophysical Journal on May 06,
2009}
\usepackage{graphicx,amssymb,amsmath,times}
\usepackage{natbib}

\begin{document}
\bibliographystyle{plainnat}
\title{Search for VHE $\gamma$-ray emission from the globular cluster M13 with the MAGIC telescope}
\shorttitle{Search of $\gamma$-rays from M13 with MAGIC} 
\shortauthors{J.
Albert et~al.}

%
\author{
H.~Anderhub\altaffilmark{a},
L.~A.~Antonelli\altaffilmark{b},
P.~Antoranz\altaffilmark{c},
M.~Backes\altaffilmark{d},
C.~Baixeras\altaffilmark{e},
S.~Balestra\altaffilmark{c},
J.~A.~Barrio\altaffilmark{c},
D.~Bastieri\altaffilmark{f},
J.~Becerra Gonz\'alez\altaffilmark{g},
J.~K.~Becker\altaffilmark{d}
% ...
}
\altaffiltext{a} {ETH Zurich, CH-8093 Switzerland}
\altaffiltext{b} {INAF National Institute for Astrophysics, I-00136 Rome, Italy}
\altaffiltext{c} {Universidad Complutense, E-28040 Madrid, Spain}
\altaffiltext{d} {Technische Universit\"at Dortmund, D-44221 Dortmund, Germany}
\altaffiltext{e} {Universitat Aut\`onoma de Barcelona, E-08193 Bellaterra, Spain}
\altaffiltext{f} {Universit\`a di Padova and INFN, I-35131 Padova, Italy}
\altaffiltext{g} {Inst. de Astrof\'{\i}sica de Canarias, E-38200 La Laguna, Tenerife, Spain}
% ...

\begin{abstract}
Based on MAGIC observations from June and July 2007, we have obtained an
integral upper limit to the VHE energy emission of the globular cluster M13  
of $F(E>200~\textrm{GeV})<5.1\times10^{-12}~\textrm{cm}^{-2}~\textrm{s}^{-1}$,
and differential upper limits for $E>140~\textrm{GeV}$. Those limits
allow us to constrain the population of millisecond pulsars within M13
and to test models for acceleration of leptons inside their magnetospheres and 
surrounding. We conclude that in M13 either millisecond pulsars are fewer than expected
or they accelerate leptons less efficiently than predicted.
\end{abstract}
\section{Introduction}
Globular clusters (GC) are very interesting sites for probing high
energy processes due to their large content of evolved objects.
Millisecond pulsars (MSP) constitute a large fraction of these objects
and it has been estimated that a typical massive GC contains of the order 
of 100 of them \citep{tavani2}. Moreover, the largest sample of MSPs 
discovered up to now in radio observations are located in the 
GC Ter 5 (23 MSP), and Tuc 47 (22 MSP) (see e.g. \citealt{camilo}). 

Fluxes of TeV $\gamma$-rays from GC detectable by current
Cherenkov telescopes have been predicted based on estimates 
on the population of MSPs and the
efficiency of lepton acceleration in their surrounding (see \citet{wlolek} and \citet{venter}). 
These $\gamma$-rays would be produced by
accelerated leptons scattering off photons of the microwave background 
radiation or the thermal emission of an extremely dense cluster of
solar mass stars inside the GC. Acceleration of leptons could take place  
in i) the shocks within the GC, coming from the collision of the winds 
of MSPs, or ii) in the pulsar inner magnetosphere or their 
wind regions. In addition, $\gamma$-rays
in the sub-TeV energy range could also be originated  in the inner MSP
magnetosphere directly, as it is predicted in the calculations by \citet{bulik,harding},
or could be produced in the vicinity of radio emitting blocked
pulsars \citep{aharonian,albert s} inside low mass binary systems \citep{tavani91}.

GCs have been observed occasionally by Cherenkov Telescopes to probe 
for this possible VHE $\gamma$-ray emission. The few experimental results 
reported in the literature are  upper limits on the emission of M13 
by the {\it WHIPPLE} Collaboration \citep{hall}, 
M15 by the {\it VERITAS} Collaboration \citep{lebohec}, and $\omega$ Centauri 
by the {\it CANGAROO} Collaboration \citep{kabuki}. Very 
recently, the {\it Fermi} LAT telescope has detected high energy 
$\gamma$-ray emission ($E>100~\textrm{MeV}$) from one of the closest and 
most massive GC, Tuc 47 
\citep{guillemot}, and {\it HESS} has obtained an upper limit of
$6.7\times10^{-13}~\textrm{ph}~\textrm{cm}^{-2}~\textrm{s}^{-1}$ for
energies $E>800~\textrm{GeV}$ \citep{hess tuc}, but given the possible complexity
of the emission in the GeV range, it is not possible to establish any
connection between these results. This {\it HESS} result constrains
the magnetic field in the pulsar nebula as a function of the
number of MSP in the GC for the model by \citet{venter}, and 
in the efficiency of the rotational energy conversion of MSPs into
relativistic leptons for the model by \citet{wlolek}.

In this paper we report the results of observations with the MAGIC
telescope of the globular cluster M13, and we discuss the constraints 
that our results impose to the population of millisecond pulsars 
and their lepton acceleration efficiency. 
M13 belongs to the class of normal globular clusters, and has an 
estimated mass of $6\times10^5~\textrm{M}_\odot$. It is located in the 
Northern sky at a distance of $~7$ kpc. Its core radius is
about $\sim1.6$ pc, with a half mass radius of $\sim3.05$ pc \citep{harris}. By now 5 
millisecond pulsars have been 
detected in M13, with periods ranging between 2 and 10 ms.
The aforementioned observation of this cluster
in search for VHE emission by the {\it WHIPPLE} Collaboration \citep{hall} 
led to a flux upper limit of 
$1.08\times 10^{-11}~\textrm{ph.}~\textrm{cm}^{-2}~\textrm{s}^{-1}$ 
at energies $E>500~\textrm{GeV}$. 
\section{Observations and data analysis}
The MAGIC telescope is an Imaging Atmospheric Cherenkov Telescope
(IACT) located at the Observatory Roque de los Muchachos on the
Canarian Island La Palma. It has an exceptional light detection efficiency
provided by the combination of a 17 m diameter mirror and 
a pixelized camera composed of 576 high quantum efficiency, hemispherical
photomultiplier tubes (PMT). This allows MAGIC to reach 
a standard trigger threshold of $\sim{60}~\textrm{GeV}$. For 
energies above $150~\textrm{GeV}$, 
angular and energy resolutions of the telescope are $\sim{0.1}~\textrm{deg.}$ and $\sim{25}$\% 
respectively (see \citealt{albert} for 
further details). Besides this, in February 
2007 its data acquisition system 
was upgraded with multiplexed 2 GHz Flash Analog-Digital converters which improved the timing resolution of the 
recorded shower images. Accordingly 
the sensitivity of MAGIC improved
significantly \citep{aliu} to 1.6\% of the Crab Nebula 
flux above $270~\textrm{GeV}$ for 50 hours of observation.

We observed M13 at zenith angles ranging from $8^{\circ}$ to $31^{\circ}$
between June 12th and July 18th of 2007 in false-source tracking (wobble) mode \citep{fomin},
with two directions at 24' distance and opposite sides of the source direction. This
technique allows for a reliable estimation of the background with no need of extra observation time.
The collected data amount to 20.7 hours after rejecting events affected by 
unstable hardware or environmental conditions. 
Besides this, events with a collected charge below 300 photo-electrons
were rejected in order to maximize the analysis sensitivity. This selection resulted  
in a sample with a peak energy of 190 GeV.

Data analysis was carried out using the standard MAGIC analysis
and reconstruction software chain, which proceeds in several steps.
Initially, a standard calibration of the PMT signal pulses
is performed  \citep{albert a}. Then, pixels containing no useful
information for the shower image reconstruction are discarded by
an image cleaning procedure \citep{aliu}.
Afterwards event image parameters are calculated \citep{hillas}  
using the surviving pixels. 
In addition to the classical Hillas parameters, 
two timing parameters are computed, namely:
the gradient of the
arrival times of the Cherenkov photons along the shower
axis; and their arrival time spread over the whole shower.
The signal-to-noise maximization is achieved using
a multidimensional classification procedure based on the
Random Forest (RF) method \citep{albert b}, where
a hadron likeness measure, the so called hadronness, is computed for each event
based on the image and time parameters. Moreover, a regression
RF trained with a Monte Carlo simulated $\gamma$-ray sample is used 
to estimate the energy on an  event by event basis. 
Finally the angle between the major axis of the 
shower image ellipse and the source position in the camera,
the so called Alpha angle, is used to select $\gamma$-ray candidates
in the direction of the source. To estimate the residual background,
the angle Alpha is also computed with respect to a position
symmetric to the source position with respect to the camera center.
In what follows, this position used to estimate the residual background 
is referred as the background region.

Main contributions to the systematic error of our analysis are
the uncertainties in the atmospheric transmission, the reflectivity
of the mirros (including losses due to surface roughness) and the light catchers,
the photon to photo-electron conversion calibration and the
photo-electron collection efficiency in the photomultiplier
front-end. A detailed
discussion of their contribution to the flux uncertainties can 
be found in \citet{albert},
where they are estimated to add up to 30\% of the measured flux value.
\section{Results}

\begin{figure}
\includegraphics[width=\linewidth]{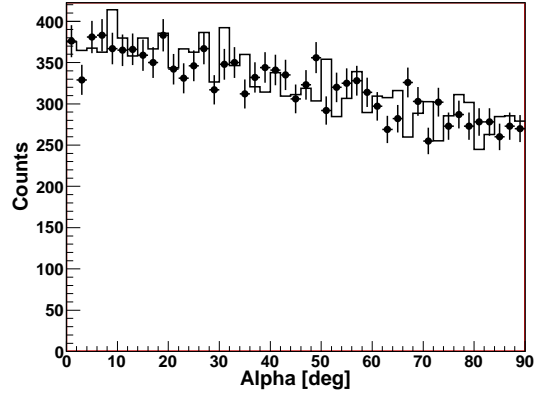}
\caption{Distribution of Alpha for the selected $\gamma$-ray candidates
from the source (black dots) and the background (histogram) regions. 
}\label{fig:alpha}
\end{figure}

Figure \ref{fig:alpha} shows the obtained distribution of the Alpha angle
for the source region and the estimated background.
It has been obtained for events surviving a hadronness cut tuned to yield  
an energy independent $\gamma$-ray selection efficiency of 80\%, 
estimated by means of a Monte Carlo simulation. 
We define the signal region as the smaller interval
in Alpha angle that contains the 80\% of the $\gamma$-rays for each 
energy bin, estimated using a Monte Carlo simulation. Their 
lower bounds are at Alpha$=0$ and the upper ones are shown in 
second column of Table \ref{tab:uldif} for each energy bin.
We find a total of $-23\pm57$ excess events after background subtraction
in this signal region for an energy range
extending from $140~\textrm{GeV}$ to $1.1~\textrm{TeV}$. In addition a search 
for signals in a region of $1~\textrm{deg}$ of radius around M13 
yields no positive detection. 
We have obtained upper limits to the VHE flux from M13 for different energy bins, shown in Table \ref{tab:uldif}.
These have been computed using 
the Rolke method by \citet{rolke} at a $95\%$ confidence level, and they take into
account a $30\%$ of systematic uncertainties in the flux level.
The upper limit to the integral flux for
energies above $E=200~\textrm{GeV}$, assuming a spectral index of 2.6, is
$5.1\times10^{-12}~\textrm{cm}^{-2}~\textrm{s}^{-1}$.

\begin{table}[h]
\caption{Differential upper limits} \vspace*{-0.3cm}
\begin{center}
\resizebox{\linewidth}{!}{
\begin{tabular}{cccccc}
\hline \hline
Energy bin  & Upper Alpha & Events & Background& Excess UL & Flux UL \\
GeV         & cut (deg)   &        & events    &(95\% CL)& (cm$^{-2}$ s$^{-1}$TeV$^{-1}$)  \\
\hline
$140~-~200$ &8 & $487$  & $517\pm23$   &37   &7.2 $\times$ 10$^{-11}$ \\
$200~-~280$ &10 & $683$  & $681\pm27$   &95   &5.1 $\times$ 10$^{-11}$\\
$280~-~400$ &8 & $254$  & $242\pm16$   &75   &2.2 $\times$ 10$^{-11}$\\
$400~-~560$ &6 & $62$   & $73\pm9$     &14   &2.4 $\times$ 10$^{-12}$\\
$560~-~790$ &4 & $32$   & $27\pm5$     &27   &2.7 $\times$ 10$^{-12}$\\
$790~-~1120$ &4 & $4$    & $5.7\pm2.4$  &5.8  &3.7 $\times$ 10$^{-13}$\\
 \hline
\end{tabular}
}
\end{center}
\label{tab:uldif}
\end{table}

\begin{figure}
\includegraphics[width=\linewidth]{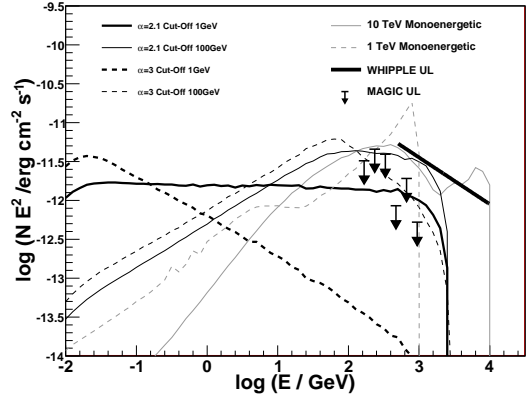}
\caption{ The MAGIC $\gamma$-ray flux upper limits for M13 compared with spectra
expected for the range of parameters of the model shown in Figure 9 and 10 of 
\citet{wlolek}. The specific $\gamma$-ray spectra are calculated for 
lepton upper energy cut-off at 3 TeV and lower energy cut-off at 1 GeV (black thick) and 100 GeV (black thin),
and power-law spectral indices of 2.1 (solid) and 3 (dashed).
The $\gamma$-ray spectra produced by mono-energetic leptons of 10 TeV and 1 TeV are shown by a grey solid curve and a dashed one respectively. All calculations are computed assuming the conservative 
value of $1$ for the free  parameter of the model $N_{\rm MSP}\cdot \eta$. 
The Whipple differential upper limit shown here has been derived
from the integral quoted in \citet{hall} assuming a spectral index of 2.6.}
\label{fig:ul}
\end{figure}

\section{Comparison with models}
\begin{figure}
\includegraphics[width=\linewidth]{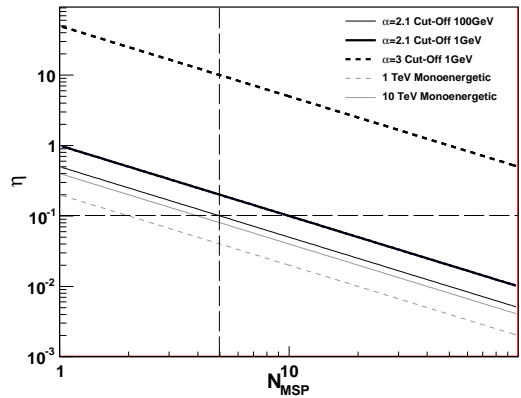}
\caption{Exclusion contours in the $N_{MSP}, \eta$ plane for the different models considered in the text. The model with parameters $\alpha$=3 and cut-off at 100 GeV overlaps the $\alpha$=2.1 with cut-off at 1 GeV one. The horizontal and vertical long-dashed black lines
show the reference values of $\eta$=0.1 and $N_{MSP}$=5 respectively.}
\label{fig:eta_n}
\end{figure}

In Figure \ref{fig:ul} we compare our flux upper limits with 
the theoretical $\gamma$-ray spectra calculated by \citet{wlolek}.
In this model, leptons are injected into the GC volume 
according to a power-law spectrum, upon acceleration in the shocks 
produced in the collisions of the pulsar winds of several MSPs. 
$\gamma$-rays are then produced via Inverse Compton scattering of photons 
from the microwave background radiation and the thermal radiation 
arising from the whole GC. Thus, the comparison of our experimental upper 
limits with the different theoretical gamma-ray spectra allows us 
to constrain the total power of injected leptons ($L_{\rm e}$). For this, 
we require the theoretical gamma-ray spectra to be lower than all 
the obtained experimental limits, we take into account the light field 
of the GC \citep{wlolek} and assume a distance of 7 pc. The upper limits to 
$L_{\rm e}$ are reported in Table 2 for different assumptions
of the spectrum of the injected leptons,
i.e. for different values of the spectral index $\alpha$ between
the minimum energy $E_{\rm min}$ and the maximum energy defined by the escape of
leptons from the shock. In the case of mono-energetic
injection of leptons we consider two different energies, 1
TeV and 10 TeV. 
Assuming characteristic values for the parameters of the MSPs in
globular clusters (surface magnetic field $10^9$ G and 
rotational period 4 ms), we can translate the limits to $L_{\rm e}$ into limits to the product of the
required number of the MSPs in M13 ($N_{\rm MSP}$) times
the efficiency of the rotational energy conversion of MSPs into 
relativistic leptons ($\eta$), shown in Table \ref{tab1}. For example, in the case of M13, 
\citet{tavani2} predicts the existence of 100 MSPs. On the other hand,
 the efficiency of lepton injection from the inner magnetospheres of
millisecond pulsars has been estimated to be $\eta\sim 0.1$ in terms 
of the extended polar gap model by \citet{muslinov}. Therefore, the 
likely value of
the product, $N_{\rm MSP}\cdot \eta$, should be of
the order of $\sim 10$. We show in the corresponding row of Table \ref{tab1} 
our estimate of the upper limit to this product for different models of injected 
spectra of leptons. Moreover, our limits in this product are at the same level 
than the recently published ones by the {\it HESS} collaboration making use
of data from 47 Tuc \citep{hess tuc}. Figure \ref{fig:eta_n} shows these limits in the $N_{MSP}, \eta$ plane, such that for each set of model parameters the area above the
corresponding curve is excluded at 95\% CL.
For most of the considered models $N_{\rm MSP}\cdot \eta$ 
is significantly below $\sim 10$. The only exception is
the model with the steep spectrum of leptons which extends down to 
1 GeV. Note that even if the 
number of MSP in M13 is only equal to 5 (as presently observed, 
\citealt{camilo}), we can already obtain the acceleration 
efficiency of leptons to be $\sim$0.1 in the case of their injection with
the hard (spectral index 2.1) and mono-energetic spectrum respectively (see Figure \ref{fig:eta_n}).

\begin{table}
\caption{Upper limits on the power of injected leptons ($L_{\rm e}$) and in $N_{\rm MSP}\cdot \eta$}
\label{tab1}
\resizebox{\linewidth}{!}{
\begin{tabular}{c c c c c c c}
\hline\hline
$E_{\rm min}$ &100 GeV&100 GeV&1 GeV&1 GeV&mono:&mono:   \\
 $\alpha$     & 2.1   & 3.0   & 2.1 & 3.0 &1 TeV&10 TeV  \\
\hline
              &       &       &     &     &     &  \\
$L_{\rm e}$   &0.6 & 1.0& 1.0& 60& 0.2& 0.5  \\
\footnotesize{($\times10^{35}~\textrm{erg}~\textrm{s}^{-1}$)}  &    &    &    &    &    &  \\
                      &    &    &    &    &    &  \\
$N_{\rm MSP}\cdot \eta$ &0.5&1.0 &  1.0 &50 &0.2&  0.4 \\
                      &    &    &    &    &    &  \\
\hline\hline
\end{tabular}
}
\end{table}

The $\gamma$-ray spectra produced in the curvature process inside the 
inner pulsar magnetospheres are predicted to cut-off below $\sim 100$ GeV
\citep{bulik,harding}. Thus, they can not extend to the 
energy region investigated by our measurement. An 
Inverse Compton $\gamma$-ray component is expected from leptons accelerated 
in the inner magnetospheres which extends $>100$ GeV. But its flux
is predicted to be at the level of several orders of magnitude below the here presented  
upper limits for most of the energy range covered by our limits. For example, 
\citet{venter} computation of VHE spectrum for GC Tuc 47 and Ter 5 predicts a flux 
level similar to the ones of the model by \citet{wlolek} for these pulsars only for 
narrow energy band above $E=1~\textrm{TeV}$.

Therefore, we conclude that the inner magnetosphere $\gamma$-ray
emission of millisecond pulsars is not likely to be detected by present
observations with an analysis threshold of the order of  $\sim 100$ GeV 
even from globular clusters containing hundreds of MSPs.

\section{Conclusions}

We have obtained the strongest upper limits to date on the VHE
$\gamma$-ray flux from the massive globular cluster M13. 
Our upper limit is $\sim 2$ times lower than the previous limit for VHE energy 
emission from M13 quoted by {\it WHIPPLE}, and 
extends to energies down to 140 GeV.
Our upper limits allow us to constrain the population of
the millisecond pulsars expected in M13 and the acceleration
scenarios of leptons by millisecond pulsars.
Our result strongly suggests that either the
number of millisecond pulsars in M13 is significantly lower than
the estimate of $\sim 100$, 
or the energy conversion
efficiency from millisecond pulsars to relativistic leptons is
significantly below the value quoted in recent modeling of high
energy processes in the magnetospheres of millisecond pulsars.
Our upper limits regarding \citet{wlolek} model parameters are the same level
than the ones obtained by the {\it HESS} collaboration making use of observations 
of Tuc 47. 
\section*{Acknowledgements}
We would like to thank the Instituto de Astrofisica de 
Canarias for the excellent working conditions at the 
Observatorio del Roque de los Muchachos in La Palma. 
The support of the German BMBF and MPG, the Italian INFN 
and Spanish MICINN is gratefully acknowledged. 
This work was also supported by ETH Research Grant 
TH 34/043, by the Polish MNiSzW Grant N N203 390834, 
and by the YIP of the Helmholtz Gemeinschaft.


%

\end{document}